\documentclass[preprint,showpacs,preprintnumbers,amsmath,amssymb]{revtex4}
\usepackage{graphicx}
\usepackage{dcolumn}
\usepackage{bm}
\usepackage{epsfig}

\def\n{{\bf{\hat{n}}}}
\newcommand{\be}{\begin{equation}}
\newcommand{\e}{\end{equation}}
\newcommand{\bear}{\begin{eqnarray}}
\newcommand{\ear}{\end{eqnarray}}

\newcommand{\begm}{\begin{pmatrix}}
\newcommand{\enm}{\end{pmatrix}}

\def\apj{ApJ}
\def\apjl{ApJL}
\def\mnras{MNRAS}
\def\prd{PRD}
\def\k{{\bf k}}

\def\x{\vec{x}}
\def\s{\tilde{r}}
\begin{document}

\title{ Gravitational  Wave Detection Using Redshifted $21$-cm Observations} 

\author{Somnath  Bharadwaj$^{1,2}$}\email{somnathb@iitkgp.ac.in} 

\author{Tapomoy Guha Sarkar$^2$}\email{tapomoy@cts.iitkgp.ernet.in} 

\affiliation{${}^1$Department of Physics and Meteorology 
I.I.T. Kharagpur, 721302, India}

\affiliation{${}^2$ Centre for  Theoretical Studies,  
I.I.T. Kharagpur, 721302, India}
  
\begin{abstract}
A gravitational wave  traversing the line of sight to a
  distant source produces a frequency shift  which contributes to
  redshift   space distortion. As a consequence,   gravitational waves  are
  imprinted   as density fluctuations  in   redshift space.   The
  gravitational wave   contribution to the  redshift space  power
  spectrum has a different $\mu$ dependence as compared to the 
  dominant contribution  from peculiar velocities. This, in principle,
  allows the two   signals to be separated. The prospect 
  of a detection is most favourable at the highest observable
  redshift $z$.  Observations of redshifted
  $21$-cm radiation from   neutral hydrogen (HI)  hold the possibility
  of   probing very high redshifts. We consider the  possibility 
  of  detecting    primordial gravitational waves using the redshift
  space   HI   power spectrum.   However,  we find
that the gravitational wave signal, though present, will not be
detectable on super-horizon scales because of cosmic variance and 
on sub-horizon scales where the signal is highly suppressed.

\end{abstract}

\maketitle

\section{Introduction}
Primordial gravitational waves are a robust prediction of
inflation \cite{grish, star}. These stochastic tensor perturbations
are generated by the same mechanism  as the  matter density
fluctuations, the ratio of the tensor perturbations  to  scalar
perturbations being quantified through the tensor-to-scalar ratio ${r}$. 
 Detecting  the  stochastic gravitational wave background is of
 considerable interest  in  cosmology  since it carries valuable
 information about the very  early universe.  The cosmological
 background of gravitational wave has  its signature imprinted on the
 CMBR temperature \cite{rubakov} and  polarization \cite{basko}
 anisotropy maps. Current CMBR observations (WMAP-$5$ Year data)
 impose an upper bound ($r < 0.43$) which is further tightened ($r <
 0.22$)  
if combined CMBR, BAO and SN data is used \cite{komatsu}.
 Detecting the gravitational
wave background is one of the important aims of upcoming PLANCK
\cite{planck} mission and future polarization based experiments 
like CMBPOL \cite{cmbpol}. 

A gravitational wave traversing the line of sight to a distant source
will contribute to its redshift in addition to that caused by 
 Hubble expansion and its peculiar velocity.
This will produce a redshift space distortion in a manner similar to
that caused by peculiar velocities \cite{kaiser}. The effect arises
due to the fact that distances are inferred from the
spectroscopically  measured redshifts. As a consequence, a
gravitational wave will manifest itself as a density fluctuation 
in redshift space. In this paper we propose this as a possible 
technique to detect the primordial gravitational wave background. 

While one could consider the possibility of detecting this at low
redshifts ($z \sim 1$) using  galaxy and quasar redshift surveys, 
we shall show that the prospects are much more favourable if the
redshift is pushed to  a value as high as possible. 

Observations of redshifted $21$-cm radiation from neutral hydrogen
(HI) can be used to measure the power spectrum of density fluctuations
at very high redshifts extending all the way to the Dark Ages 
($30 < z <200$)
\cite{zaldaloeb}. Redshift space distortions make an important
contribution  to this signal \cite{bhaali}. We investigate the
possibility of using this to detect primordial gravitational waves. 
We note that the imprint of gravitational waves on the 21-cm signal
from the Dark Ages has also been considered  in an ealier work
\cite{lewis}. 

\section{Formulation}    
The radial component of peculiar velocity  introduces  a  redshift 
$z_v =v/c$  in excess of the cosmological redshift which 
arises due to the expansion of the universe. This distorts our view of
the matter distribution in the three dimensional redshift space,
where the radial distance is inferred from the measured redshift.  
As a consequence the density contrast $\delta = \delta \rho/\rho$
measured in redshift space $\delta^s$ is different from  the actual
density contrast $\delta^r$, and \cite{hamilton}
\be 
\delta^s=\delta^r-\frac{c}{aH(a)} \frac{\partial z_v}{\partial x}
\label{eq:a1}
\e
where $H(a)$ is the Hubble parameter and $x$ the comoving distance to
the source. 
We see that any coherent velocity pattern (in-fall or outflow)
manifests itself 
as a density fluctuation in redshift space. This  takes a 
particularly convenient form in Fourier space if we 
assume that the peculiar velocities  are  produced by the density
fluctuations $\delta^r$.   We then have 
\be
\Delta^s(\k)=(1 + f \mu ^2) \Delta^r(\k)
\e
where  $\Delta^s$ and $\Delta^r$ are the Fourier transforms of
$\delta^s$ and $\delta^r$ respectively, 
$f =d \ln D/d \ln a \approx \Omega_m^{0.6}$, $D$ being the
growing mode of density perturbations  and 
$\mu=\hat{n} \cdot \k/k$ is the cosine of the angle between the line
of sight $\n$ and the wave vector $\k$.  It follows that 
the power spectrum of density
fluctuations in redshift space $P^s(k)$, is related to its real space 
counterpart  $P^r(k)$ as \cite{kaiser}
\be 
P^s(\k)=(1 + f \mu ^2)^2 P^r(\k)
\e

 A gravitational wave $h_{ab}(\x,\eta)$ which is 
a tensor metric perturbation 
\be
ds^2 = a^2 \, \left[ c^2 \, d \eta^2 - (\delta_{ab} + h_{ab}) 
dx^a \, dx^b \right]
\e
makes an additional contribution \cite{mukhanov}
\be
z_h =  \frac{1}{2} n^a n^b \,
\int_{\eta_e}^{\eta_0} 
 h^{'}_{ab}(\x, \eta) \, d \eta \,
\e
to the redshift 
along the line of sight of the unit vector  $\hat{n}$. Here prime
 denotes a partial derivative with respect to $\eta$, $\eta_e$ and
 $\eta_0$ respectively refer to the photon being emitted and the 
present epoch when the photon is observed, and $\x=\hat{n}
 (\eta_0-\eta)$ is the photon's spatial trajectory. Considering $z_h$ the 
gravitational wave contribution  to the redshift, we have an  additional
contribution 
\be 
\delta^s_h=-\frac{c}{aH} \frac{\partial z_h}{\partial x}
\e
to $\delta^s$ the density contrast in redshift space (eq. \ref{eq:a1}).
Simplifying this using $x = c(\eta_{0} - \eta_e)$ we have 
\be
\delta^s_h=\frac{1}{2aH} n^a n^b \,
  h^{'}_{ab}
\label{eq:b1}
\e

We consider the primordial gravitational waves which 
we expand  in Fourier modes as
\be
h_{ab}(\x,\eta)=\int \tilde{h}_{ab}(\k,\eta) e^{i \k \cdot \x}
\frac{d^3 \k}{(2 \pi)^3}
\e
and decompose  $\tilde{h}_{ab}(\k,\eta)$ in terms of the two polarization
tensors $e^{+}_{ab}$ and $e^{\times}_{ab}$ as \cite{tarunbharad}
\be
\tilde{h}_{ab}(\k, \eta) 
= h(k, \eta) \left [ e^{+}_{ab} a^{+}(\k) +
  e^{\times}_{ab} a^{\times}(\k) \right] \frac{\sqrt{(2 \pi)^3 P_h(k)}}{2}\,.
\e
Here $h(k,\eta)$ quantifies the temporal evolution, and $h(k,\eta)=3
j_1(k c \eta)/(k c \eta)$ in a matter dominated universe, $j_1$ being
the spherical Bessel function of order unity. The polarization tensors
are normalized to $e^{+}_{ab}  e^{+ \, ab} =
e^{\times}_{ab} e^{\times \, ab} =2$ and $ e^{+}_{ab}
e^{\times \, ab}=0 $, $P_h(k)$ is the primordial
gravitational wave   power spectrum \cite{komatsu}
  and $a^{\times}(\k), a^{+}(\k)$
are Gaussian random  variables such that 
 \be
\langle \tilde{h}^*_{ab}(\k, \eta) \tilde{h}^{ab}(\k^{'}, \eta)
\rangle 
 = (2 \pi)^3 \delta^3(\k-\k^{'}) h^2(k,\eta) P_h(k)
\e

Let us first consider a single Fourier mode of  gravitational wave
with $\k$ along the $z$ direction,  and represent the line of sight as     
\be
\hat{n}=\sin \theta (\cos \phi \, \hat{i} +\sin \phi \, \hat{j})+ 
\cos \theta \, \hat{k}\,.
\e
We can than  express  eq. (\ref{eq:b1}) as 
\be 
\Delta^s(\k,\eta)=\frac{h^{'}}{4aH} \sin^2 \theta \left[ \cos 2 \phi \,
  a^+(\k) + \sin 2 \phi \, a^{\times}(\k) \right ]
 \sqrt{(2 \pi)^3   P_h(k)}\,. 
\e
This can be equivalently interpreted with $\hat{n}$ fixed and the
direction of $\k$ varying. We use this to calculate $P^s_h(\k)$ 
the gravitational wave contribution to the power spectrum of density
fluctuations in redshift space  
\be
P^s_h(\k)=\sin^4 \theta  \,\left\{  \left[ \frac{h^{'}}{4aH} \right]^2 \, \,
P_h(k)  \right\}
\label{eq:c1}
\e
Thus the  total power spectrum of density fluctuations in redshift
space is  
\be 
P^s(\k)=(1 + f \mu ^2)^2  P^r(k) + {(1-\mu^2)}^2 P^r_h(k)
\label{eq:d1}
\e
where $P^r_h(k)$ refers to the terms in $\{\, \,  \}$ in
eq. (\ref{eq:c1}). Here  $P^r(k)$ and $P^r_h(k)$ are respectively 
the matter and  gravitational wave contributions to the power spectrum
of density fluctuations in redshift space. Both  $P^r(k)$ and
$P^r_h(k)$ are  to be evaluated at the epoch corresponding to the
redshift under  observation.    

The contributions from $P^r(k)$ and $P^r_h(k)$ have different $\mu$
dependence. This, in principle, can be used to separately estimate 
the gravitational wave and the matter contributions from  the observed 
redshift space power spectrum. While the matter 
contribution is maximum when $\k$ and $\hat{n}$ are parallel, the 
gravitational wave contribution peaks when the two are mutually
perpendicular.

\section{Results}

\begin{figure}
\begin{center}
\mbox{\epsfig{file=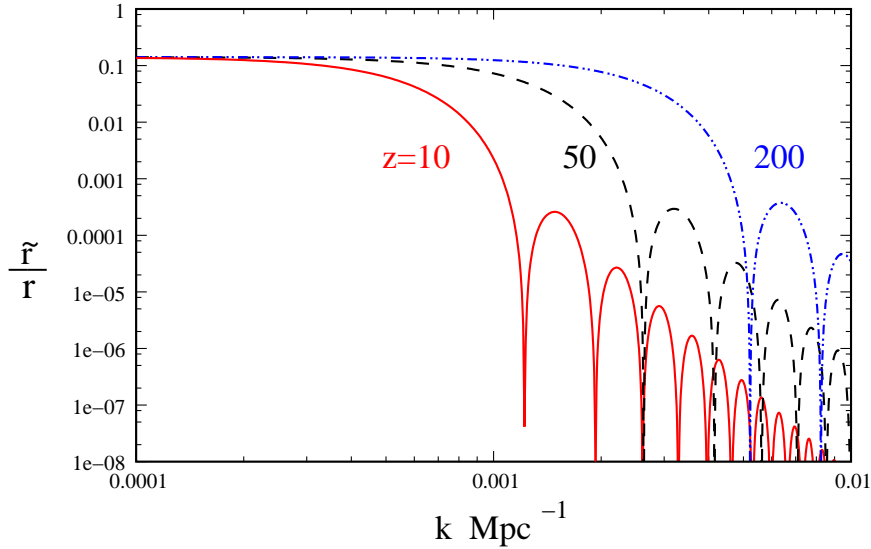,width=0.70\textwidth,angle=0}}
\caption{This shows the ratio $\s/r$ at different $z$. This is
  predicted to have a constant value $\sim 0.16$ on super-horizon
  scales in the $\Omega_m=0.3$ LCDM model considered here.}
\label{fig:1}
\end{center}
\end{figure}

We use $\s=P^r_h(k)/P^r(k)$  to quantify the ratio of 
 tensor  perturbations to scalar perturbations in the
redshift space power spectrum. Assuming $n_s=1$, $n_T\ll1$,  
the value of $\s$ is constant on super-horizon scales $(k c \eta
\ll1)$. This value is $\s = r/4$ if  $\Omega_m = 1$, 
and somewhat smaller (Figure  \ref{fig:1}) with $\s = 0.16 r$ for 
 $\Omega_m=0.3$  in the LCDM  model.  Gravitational waves decay inside
 the horizon  whereas matter perturbations grow on these scales. The
 ratio $\s(k)$ is oscillatory and is severely suppressed on sub-horizon scales $(k c \eta \ll 1)$.

The prospect of detecting the gravitational wave signal is most
favourable on super-horizon scales $(k \le k_{\rm H}=(c \eta)^{-1})$.
The $k$ range  amenable for such observations (Figure \ref{fig:1})
increases with redshift $z$ (smaller horizon $c \eta$).  Observations
of redshifted $21$-cm radiation  hold the potential of measuring  the
redshift space power spectrum in the $z$ range
$(30-200)$\cite{zaldaloeb, bhaali}, where the pre-reionization HI  
 signal will be seen in absorption against the CMBR. 
Gravitational waves will make a $\sim r \times 16 \%$ contribution to
the HI signal on scales $k \le k_{\rm H}$.

\section{Feasibility of Detection}
The cosmological HI signal will be buried in foregrounds \cite{bhaali1,
  santos, mcquinn, datta,ali}   
which are
expected to be orders of magnitude larger than the signal. The
foregrounds are  continuum sources whose spectra are expected to be
correlated over large frequency separations, whereas the   HI signal,  
a line emission, is expected to be uncorrelated  beyond a frequency
separation.  While this, in principle, can be used to separate the HI
signal from the foregrounds, it should be noted that the frequency
separation beyond which the HI signal becomes uncorrelated increases
with $z$ and angular scale. This is a potential problem for the
detection of the gravitational wave signal. In the subsequent
discussion we have assumed that the foregrounds have been removed from
the HI signal. 

The distinctly different $\mu$  dependence of the scalar and
gravitational wave components of the 
redshift space power spectrum can in principle be used to separate the
two signals. Expressing the $\mu$ dependence   \cite{barlo}   as
$P^s(k, \mu) = P_0(k) + P_2(k) \mu^2 + P_4(k) \mu^4$,
the gravitational wave component can be estimated using 
$P_h^r(k) = [ P_0(k) - P_2(k) ]/2$.
For a cosmic variance limited experiment, the error in $P_2(k)$ and
$P_0(k)$  would be  $ \delta P(k)/ P(k) \sim 1/\sqrt{N(k)}$
\cite{seo,Mao,mcquinn,low,pritch}, where $N(k)$ denotes the number of
$\k$ modes within  
the comoving volume of the survey. Thus $ N(k) > \s^{-2} \sim 10^4$
modes would be needed for a  detection of the gravitational
wave signal.

The number of modes with a comoving wave number between $k$ and $k
+dk $ is $dN(k) = k^2 \ dk \mathcal{V}/{(2\pi)}^2$, where
$\mathcal{V} $ is the comoving survey volume. Assuming a survey
between $ z=20 $ to $ z= 200$, and using a $k$ bin $dk = k/10$, 
we have $N(k)=10$  for $k=k_{\rm H} \sim 0.002 \ {\rm Mpc}^{-1}$.

It is, in principle, possible to carry out HI observations in the
entire $z$ range $z=0$ to $z=200$ \cite{bhaali1} and thereby increase
the volume.  Of the entire survey volume
$\mathcal{V}_0$, for a mode $k$  only a volume   $\mathcal{V}(k) = 
\mathcal{V}_0 -(4 \pi/3)(c \eta_0-k^{-1})^3 $      where the mode is
super-horizon contributes to  the  signal.  Further, the largest mode
$k_{max}$ is the one that entered the horizon at $z=200$, and the smallest
mode $k_{min}$ has wavelength comparable to the radius of the survey
volume.  We then have, assuming a full sky survey,  
\be
N = (2 \pi^2)^{-1} \int_{k_{min}}^{k_{max}}  
\ \mathcal{V}(k) k^2 dk   
\e
which gives $N \sim 100$. The number of independent modes is too small
for a  measurement at a level of precision that will allow the
gravitational wave component to be detected. In conclusion, we note
that the gravitational wave signal, though present, will not be
detectable on super-horizon scales because of cosmic variance and 
on sub-horizon scales where the signal is highly suppressed.

\end{document}